\documentclass[]{spie}
\usepackage{graphicx}
\usepackage{dcolumn}
\usepackage{bm}
\usepackage{mathbbol}

\RequirePackage{amsmath}
\RequirePackage{amssymb}

\providecommand\onlinecite[1]{\citen{#1}}

\newcommand{\mathcommand}[3][0]{\newcommand{#2}[#1]{\ensuremath{#3}}}

\newcommand{\ts}[2]{{#1}_{\textnormal{#2}}} 
\newcommand{\tsc}[2]{{#1}_{\textsc{#2}}} 

\newcommand{\abbrev}[1]{{\scshape\lowercase{#1}}}  

\renewcommand{\vec}[1]{\ensuremath{\mathbf{#1}}}

\newcommand{\be}{\begin{equation}}
\newcommand{\ee}{\end{equation}}
\newcommand{\splitstart}{\begin{equation}\begin{split}}
\newcommand{\splitstop}{\end{split}\end{equation}} 

\newcommand{\refeq}[1]{Eq.~\eqref{#1}}

\newcommand{\reffig}[1]{Fig.~\ref{#1}}

\newcommand{\ham}[1][]{\ensuremath{{H}_{\text{#1}}}} 

\mathcommand[1]{\smallket}{|#1\rangle}
\mathcommand[1]{\smallbra}{\langle #1|}
\mathcommand[1]{\bigket}{\bigl|#1\bigr\rangle}
\mathcommand[1]{\bigbra}{\bigl\langle #1\bigr|}
\mathcommand[1]{\biggket}{\biggl|#1\biggr\rangle}
\mathcommand[1]{\biggbra}{\biggl\langle #1\biggr|}
\mathcommand[1]{\ket}{\left| #1 \right\rangle}  
\mathcommand[1]{\lket}{\bigl| #1 \bigr)}  
\mathcommand[1]{\bra}{\left\langle #1 \right|}  
\mathcommand[1]{\lbra}{\bigl( #1 \bigr|}  

\mathcommand{\te}{\text{e}}
\mathcommand{\vactext}{\text{vac}}
\mathcommand{\vacket}{\ket\vactext}
   \mathcommand{\es}{\langle S(\Omega) \rangle}
    \mathcommand{\Sb}{\ts\es{before}}
    \mathcommand{\Sa}{\ts\es{after}}

\mathcommand\mub{\tsc\mu{b}}

\mathcommand\GHz{\text{~GHz}}
\mathcommand\ms{\text{~ms}}

\newcommand{\QD}{\abbrev{QD}}
\newcommand{\QDs}{\abbrev{QD}s}
\newcommand{\FID}{\abbrev{FID}}
\newcommand{\CW}{\abbrev{CW}}

\newcommand{\pdf}{\abbrev{pdf}}
\mathcommand{\am}{\langle m(\vec{r})\rangle}

\title{Nuclear Feedback in a Single Electron-Charged Quantum Dot under Pulsed Optical Control}

\newcommand\ginzton{E. L. Ginzton Laboratory,
        Stanford University,
        Stanford, California 94305, USA}
\newcommand\NII{National Institute of Informatics,
        Hitotsubashi 2-1-2, Chiyoda-ku,
        Tokyo 101-8403, Japan}
\newcommand\wurzburg{Technische Physik, Physikalisches Institut,
        Wilhelm Conrad R\"{o}ntgen Research Center for Complex Material Systems,
        Universit\"{a}t W\"{u}rzburg,
        Am Hubland, D-97074 W\"{u}rzburg, Germany}
\newcommand\HRL{HRL Laboratories, LLC, 3011 Malibu Canyon Rd., Malibu, CA 90265, USA}

\author{Thaddeus~D.~Ladd\supit{a,b,}\!
        \footnote{~~Present address: \HRL}~,
        David~Press\supit{a},
        Kristiaan~De~Greve\supit{a},
        Peter~L.~McMahon\supit{a},
        Benedikt~Friess\supit{a,c},
        Christian~Schneider\supit{c},
        Martin~Kamp\supit{c},
        Sven~H\"{o}fling\supit{a,c},
        Alfred~Forchel\supit{c},
        and
        Yoshihisa~Yamamoto\supit{a,b}
\skiplinehalf
\supit{a}\ginzton\\
\supit{b}\NII\\
\supit{c}\wurzburg
}

\begin{document}
\maketitle
\begin{abstract}
Electron spins in quantum dots under coherent control exhibit a number of novel feedback processes.  Here, we present experimental and theoretical evidence of a feedback process between nuclear spins and a single electron spin in a single charged InAs quantum dot, controlled by the coherently modified probability of exciting a trion state. We present a mathematical model describing competition between
optical nuclear pumping and nuclear spin-diffusion inside the
quantum dot.  The model correctly postdicts the observation of a hysteretic sawtooth pattern in the free-induction-decay of the single electron spin, hysteresis while scanning a narrowband laser through the quantum dot's optical resonance frequency, and non-sinusoidal fringes in the spin echo.  Both the coherent electron-spin rotations, implemented with off-resonant ultrafast laser pulses, and the resonant narrow-band optical pumping for spin initialization interspersed between ultrafast pulses, play a role in the observed behavior. This effect allows dynamic tuning of the electron Larmor frequency to a value determined by the pulse timing, potentially allowing more complex coherent control operations.
\end{abstract}
\keywords{Quantum dot, nuclear polarization, ultrafast control}

\section{Introduction}
The control of a single electron spin in a semiconductor quantum dot (\QD) using single, ultrafast optical pulses has emerged as a promising path towards the development of high-speed, optically driven quantum information processors. One remaining obstacle for using \QD\ spins as qubits is the need to compensate for the drift of random nuclear fields.  This problem may be alleviated by one of several recently discovered nuclear feedback phenomena observed in \QDs\ under coherent control.  Such phenomena occur in electrically
controlled double \QDs, in which transition processes between electron singlet and triplet states allow the manipulation of interdot nuclear spin polarizations, improving coherent
control~\cite{vandersypen_singlet-triplet,Reilly_Zamboni_Science,new_yacoby}.
In single \QDs\ under microwave control, nuclear effects
dynamically tune the electron spin resonance to the applied
microwave frequency~\cite{delftlock}.  Tuning effects are also
observed in two-color continuous-wave (\CW) laser experiments in self-assembled \QDs, in which the appearance of coherent electronic effects such as population trapping are modified by nonlinear feedback processes with nuclear spins~\cite{sham_cpt,imamoglu_cpt}. Finally, nuclear spins have been shown to dynamically bring ensembles of inhomogeneous \QDs\ into spin-resonance with a train of ultrafast pulses~\cite{greilich_nuclear,Reinecke_nuclear}.

Recently, similar nuclear feedback phenomena have also been observed for a single self-assembled \QD\ under pulsed ultrafast control\cite{PRLpaper}.  The
effect is evident when measuring the familiar ``free-induction
decay" (\FID), equivalent in this context to a Ramsey
interference experiment, of a single spin in a single \QD\
under pulsed control.  The Larmor frequency of the electron
spin is dynamically altered by the hyperfine interaction with
\QD\ nuclei; the nuclear polarization is in turn altered by the
measurement results of the \FID\ experiment. The result is a
feedback loop in which the nuclear hyperfine field stabilizes
to a value determined by the timing of the pulse sequence.  The experimental observation is treated in more detail in Ref.~\onlinecite{PRLpaper}.  In
what follows, we elaborate on a mathematical model to explain the effect.  We also show how this model can be used to describe related hysteretic optical pumping phenomena and non-sinusoidal spin-echo fringes.

These observations and this model could have strong impact on the viability of optically controlled \QDs\ for quantum information processing.  Prior to the discovery of nuclear control effects, nuclear spins were generally regarded as a problem for quantum information processors, since they cause adverse effects such as inhomogeneous broadening and non-Markovian decoherence processes. While useful roles such as long-lived quantum memory were envisioned~\cite{tml03}, their practicality was questionable due to limited amounts of nuclear polarization and memory loss due to nuclear spin diffusion.  The present results, along with similar work~\cite{vandersypen_singlet-triplet,Reilly_Zamboni_Science,delftlock,new_yacoby,sham_cpt,imamoglu_cpt,greilich_nuclear,Reinecke_nuclear}, show that nuclear spins may provide novel methods for the dynamic \emph{tuning} and \emph{locking} of electron spin resonances for electrons trapped in \QDs.  This may benefit the stability of future coherent devices based on \QDs.

In Sec.~\ref{sec:experiment}, we briefly describe the experimental circumstances and, and in Sec.~\ref{results}, we show how nuclear feedback appeared in observations. Section~\ref{physicalprocesses} then describes the physical processes which we believe are most likely to account for the observations.  With these assumed, we present in Sec.~\ref{themodel} a simple mathematical model that replicates the experimental observations.  We then conclude with some prospects for how this work may impact quantum information processing.

\section{Experiment}

\label{sec:experiment}
The experimental apparatus is similar to that described in
Refs.~\onlinecite{pressecho} and \onlinecite{pressnature}; we refer the reader to these references for more details of the experiment, including apparatus schematics.

\subsection{Quantum Dot Sample}
\begin{figure}[t]
\begin{center}
\includegraphics[width=0.5\columnwidth]{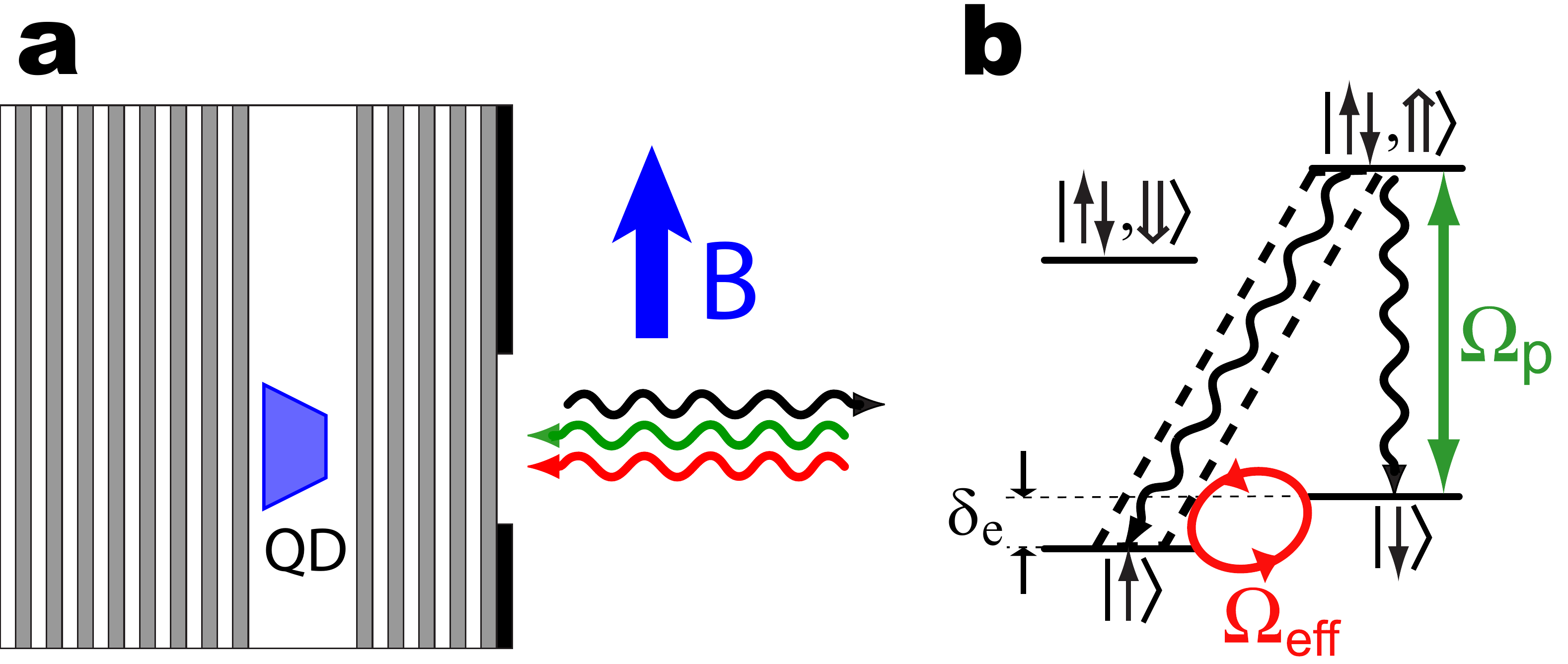}
\end{center}
\caption{
(a) Sample geometry; a single QD is embedded between AlAs/GaAs
mirrors; a transverse magnetic field $B$ is applied.
(b) Level structure and optical interactions used in the
experiment.  $\Omega_p$ (green) labels the CW optical pump,
$\ts\Omega{eff}$ (red) labels the pulsed coherent rotation.  Wavy lines indicate
spontaneous emission; the photons collected for measurement are indicated
by the dashed box.}
\label{sample}
\end{figure}

The sample is grown by molecular beam epitaxy to contain roughly $2 \times
10^{9}\text{~cm}^{-2}$ self-assembled InAs \QDs, situated
roughly 10~nm above a $\delta$-doping layer of $\sim 4 \times
10^{9}\text{~cm}^{-2}$~Si donors.  About one-third of the \QDs\
were negatively charged by the $\delta$-doping layer, and a
single charged \QD\ was spectroscopically isolated at around
940~nm for the experiments.  As shown in
Fig.~\ref{sample}(a), the \QDs\ were embedded at the center
of a planar GaAs microcavity ($Q ~\sim 200$) with 24 and 5
pairs of AlAs/GaAs $\lambda/4$ layers in the bottom and top
mirrors, respectively, which served to direct the \QD\ emission
towards the collection lens and decrease the optical power
required for spin rotation.  As shown in Fig.~\ref{sample}(b), the two spin states of the \QD\ electron, $\left|\downarrow \right\rangle$ and $\left|\uparrow \right\rangle$, are split with a Larmor frequency $\delta_{\text{e}} = g_{\text{e}} \mu_{\text{B}}
B_{\text{ext}}/\hbar$.  Except where noted, all data presented here uses a $\ts{B}{ext}=4$~T magnetic field, which leads to $\ts\delta{e}/2\pi= 25.3$~GHz,   The field is applied in Voigt geometry, perpendicular to the optical axis and sample growth direction; see Fig.~\ref{sample}(a).  In this geometry, the electron-spin ground states each couple optically to both heavy-hole trion states $\left|\downarrow \uparrow, \Downarrow \right\rangle $ and $\left|\downarrow \uparrow, \Uparrow \right\rangle $.  These trion states each contain two electrons in a spin-singlet and an unpaired heavy hole.

\subsection{Ultrafast Coherent Control}
At the core of our experiment is the coherent control of a single electron spin using ultrafast pulses.  We
coherently manipulate the electron spin state by applying to
the \QD\ a circularly-polarized, 4-ps optical pulse that is
detuned by $\sim 150$~GHz below the exciton transitions.  The
pulse rotates the spin approximately around the optical axis, allowing superpositions of $\left|\downarrow \right\rangle$
and $\left|\uparrow
\right\rangle$~\cite{fast_rotations_prop,pressnature,pressecho,awschsingle}.

There are two equivalent points of view as to how the pulse accomplishes a coherent rotation.  One point of view, described in Refs.~\onlinecite{fast_rotations_prop,pressnature} and elsewhere, is that the pulses work via stimulated Raman transitions.  Physically, a train of pulses from a modelocked laser may be considered to have a comb of frequencies; two peaks of such a comb can provide the two elements of a Raman transition.  The train of pulses (in the time domain) which leads to the frequency comb (in the frequency domain) will, if their timing is tuned to the Larmor period of the single electron spin, always rotate the electron spin with the same phase, constructively achieving a rotation.  If the pulse power or oscillator strength is high enough, as it is in our case, substantial rotations can occur with just a single pulse, and the relative timing between just a pair of pulses can achieve complete qubit control~\cite{pressnature}.  Mathematically, the action of the pulse may be derived using adiabatic elimination of the excited states, in which time-derivatives of excited states in either the Schr\"odinger or master equation are set to zero.  As a result, the single-pulse rotation angle is approximated as
\be
\label{theta_eq}
\theta \approx \int \frac{\Omega^2(t)}{\Delta}dt,
\ee
where $\Delta$ is the detuning and $\Omega(t)$ is the time-varying optical Rabi frequency, proportional to the pulse's electric field.

The mathematical procedure of adiabatic elimination includes two approximations: the adiabatic approximation that, due to the large detuning, the system remains in a ``slowly" changing ground state; and the approximation inherent in first-order perturbation theory.  Only the first approximation is used in the point of view for ultrafast coherent control as an AC-stark shift, indicated in Ref.~\onlinecite{awschsingle} and elsewhere.  For this picture, we note that the pulse length is much shorter than the Larmor precession period.  On this timescale, the electron ground state is effectively degenerate, and as a result there exists an optically dark ground state (for example, $\ket{\uparrow}-\ket{\downarrow}$, neglecting normalization), and an orthogonal optically bright state (for example, $\ket{\uparrow}+\ket{\downarrow}$).  The details of these states will depend on the selection rules of the dot, which in turn depends on details of the dot's symmetry.  In this point of view, the arrival of the pulse ``slowly" mixes the bright state and the excited state to which it is optically connected.  Because of the large detuning, the adiabatically changing mixed state reduces in energy by
\be
\label{delta_omega_eq}
\delta\omega(t)=\frac{\Delta}{2}\left[\sqrt{1+\frac{4\Omega^2(t)}{\Delta^2}}-1\right].
\ee
This is the AC-stark shift of the bright state.  A spin beginning in a superposition of bright and dark states will therefore see its bright state component shifted by the time integral of $\delta\omega(t)$, thereby accomplishing a spin rotation.  Note that the first-order approximation of $\int \delta\omega(t) dt$ in the limit $\Delta\gg\Omega(t)$ provides the rotation angle obtained from the adiabatic elimination point of view, \refeq{theta_eq}.  Also note that the nonlinear dependence on the rotation angle $\theta$ with pulse power, as reported in Ref.~\onlinecite{pressnature}, may be described by the time-integral of \refeq{delta_omega_eq}.

\subsection{Pulse Sequences}
\label{pulsesequence}
Before being manipulated by a sequence of rotation pulses, the
spin is first initialized into the ground state $\left|\uparrow
\right\rangle$ by optical pumping [\reffig{sample}(b)].
A pulse of duration $\ts{T}{p}=26$~ns from a narrow-band laser, gated with an
electro-optic modulator (\abbrev{EOM}), drives the $\left|\downarrow
\right\rangle\leftrightarrow \left|\downarrow \uparrow ,
\Uparrow \right\rangle$ transition at rate
$\Omega_\text{p}$. Spontaneous decay then shelves the spin into
$\left|\uparrow \right\rangle$ within a nanosecond timescale.
The optical pumping step also serves to measure the spin-state
resulting from the previous rotation sequence: if the spin was
rotated to $\left|\downarrow \right\rangle$ then a single
photon will be emitted from the $\left|\downarrow \uparrow ,
\Uparrow \right\rangle \rightarrow\left|\uparrow
\right\rangle$ transition as the spin is re-initialized.  This
photon is spectrally filtered by a double-monochromator and
detected using a single-photon counter.

\begin{figure}[t]
\begin{center}
\includegraphics[width=0.7\columnwidth]{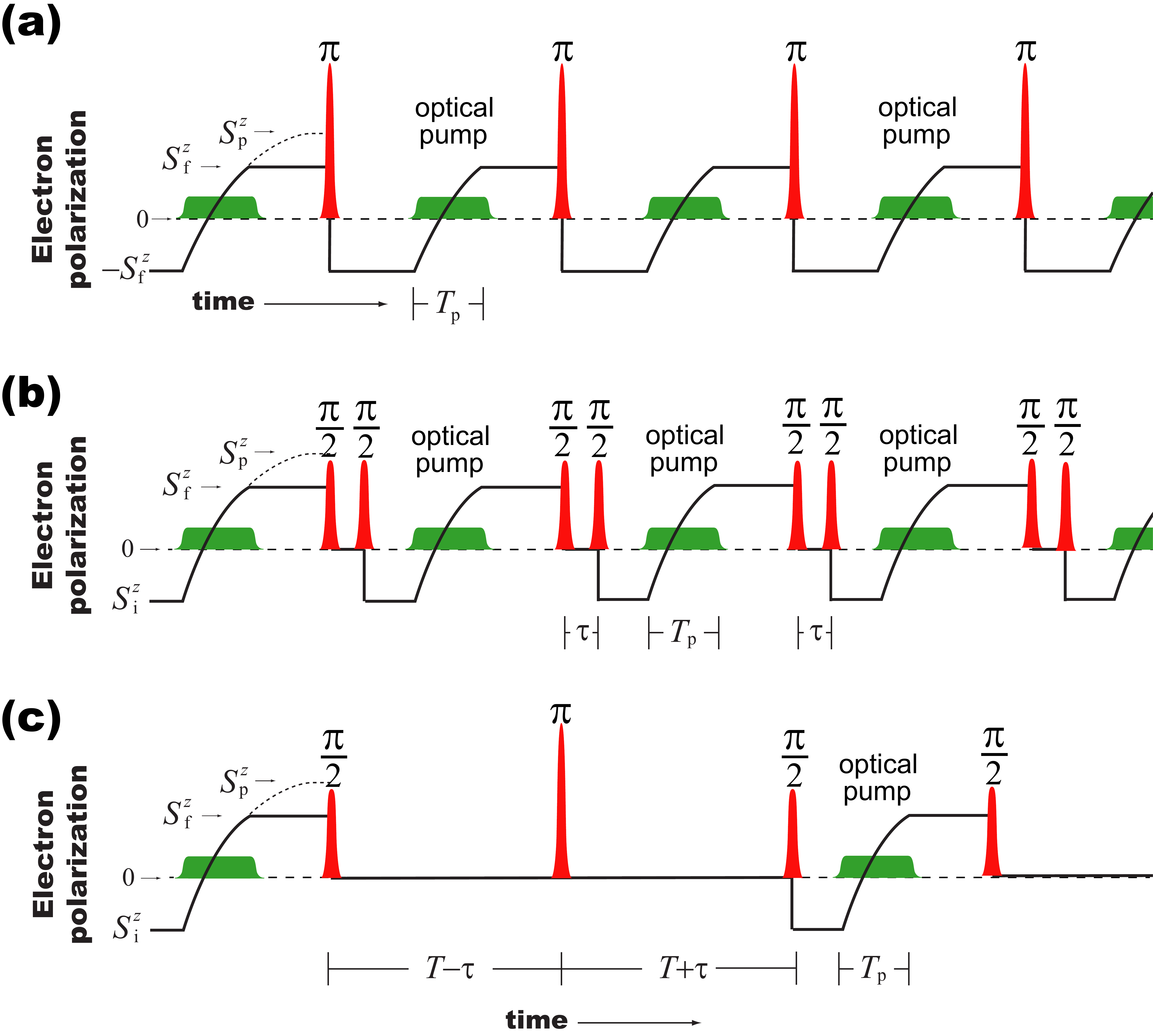}
\end{center}
\caption{
The three pulse sequences employed.
(a) One-pulse, optical pumping sequence.  After the optical pump (green pulse) polarizes the spin to final polarization $\ts{S}{f}^z$, as shown by the black line, an ultrafast $\pi$ pulse (red) inverts the polarization, and the sequence is repeated.  The polarization $\ts{S}{f}^z$ depends on the wavelength $\lambda$ of the pumping laser, which allows an absorption spectrum.
(b) Two-pulse, \FID\ or Ramsey interference sequence.  The optical pump (green pulse) polarizes the spin from $\ts{S}{i}^z$ to a final polarization $\ts{S}{f}^z$, after which it is rotated to the Bloch-sphere equator by an ultrafast $\pi/2$ pulse (red).  A second $\pi/2$ pulse a time $\tau$ later rotates the spin to a location on the Bloch sphere with $z$-projection $\ts{S}{i}^z$ that depends on $\tau$, and the sequence is repeated.  During optical pumping, $\ts{S}{f}^z-\ts{S}{i}^z$ is measured.
(c) Three-pulse, spin-echo sequence.  The sequence is the same as (b), except an optical $\pi$-pulse is inserted approximately halfway between the $\pi/2$ pulses, refocusing nuclear Overhauser shifts in the electron spin's precession.  The $z$-projection of the spin after the last pulse, $\ts{S}{i}^z$, depends on the offset $\tau$ of the $\pi$ pulse from the center of the $\pi/2$ pulses.
}
\label{sequences}
\end{figure}

In this work we describe the results of three different pulse sequences, summarized in \reffig{sequences}.

\subsubsection{One-pulse Sequence: Optical Pumping}
  In a one-pulse or ``optical pumping" sequence, illustrated in \reffig{sequences}(a), a single ultrafast $\pi$ pulse is applied; this pulse is also gated by an \abbrev{EOM}.   This sequence is designed to give maximal signal during optical pumping, allowing measurements such as the spectral absorption while scanning the pumping laser across the resonance.

\subsubsection{Two-pulse Sequence: Ramsey interferometry/Free Induction Decay}
  In a two pulse sequence, illustrated in \reffig{sequences}(b), the spin is first initialized into $\left|\uparrow \right\rangle$.  After the first $\pi/2$ rotation, the electron spin freely precesses around the equator of the Bloch sphere at its Larmor precession frequency, including any Overhauser shifts.   A second $\pi/2$ rotation introduces a projection onto the $z$-axis which varies sinusoidally with $\tau$; that projection is measured during the subsequent initialization step as the entire sequence is repeated.  After averaging, the resulting trion count rate is expected to show sinusoidal fringes as a function of $\tau$.

    In practice, the two timed pulses are achieved by first splitting optical pulse with an unbalanced Mach-Zender interferometer and then gating single pulses with an \abbrev{EOM}.  A retroreflector on a computer-controlled stage allows the time delay $\tau$ between the two rotation pulses to be scanned by up to 300~ps.

\subsubsection{Three-pulse Sequence: Spin-Echo}
    In the spin-echo experiment, illustrated in \reffig{sequences}(c), the three separate pulses from the mode-locked laser are each gated by multiple \abbrev{EOM}s on two split optical paths.  The two paths have different optical attenuations, assuring a $\pi/2-\pi-\pi/2$ Hahn echo sequence.  We denote the time between the first $\pi/2$ rotation and the $\pi$ rotation $T$, which is necessarily an integer multiple of the pulse period of the mode-locked laser, 13~ns.  However, the optical path for the middle $\pi$-pulse is sent through a separate optical path with variable length tuned by the computer-controlled stage and retroreflector, allowing $\pi/2$-to-$\pi$-pulse separation to be modulated to $T-\tau$ while the $\pi$-to-$\pi/2$-pulse separation is $T+\tau$.  Fringes are therefore observed as a function of $\tau$ at twice the Larmor frequency.  The decay of these fringes as a function of $T$ allows a measurement of $T_2$, described elsewhere~\cite{pressecho}.  Here, the focus is on the behavior with respect to $\tau$.

\section{Observations}
\label{results}
    Nuclear feedback effects are observed during the application of all three sequences, but they manifest in different ways during each sequence, as we now describe.
\subsection{Optical Pumping}
\begin{figure}
\begin{center}
\includegraphics[width=0.5\columnwidth]{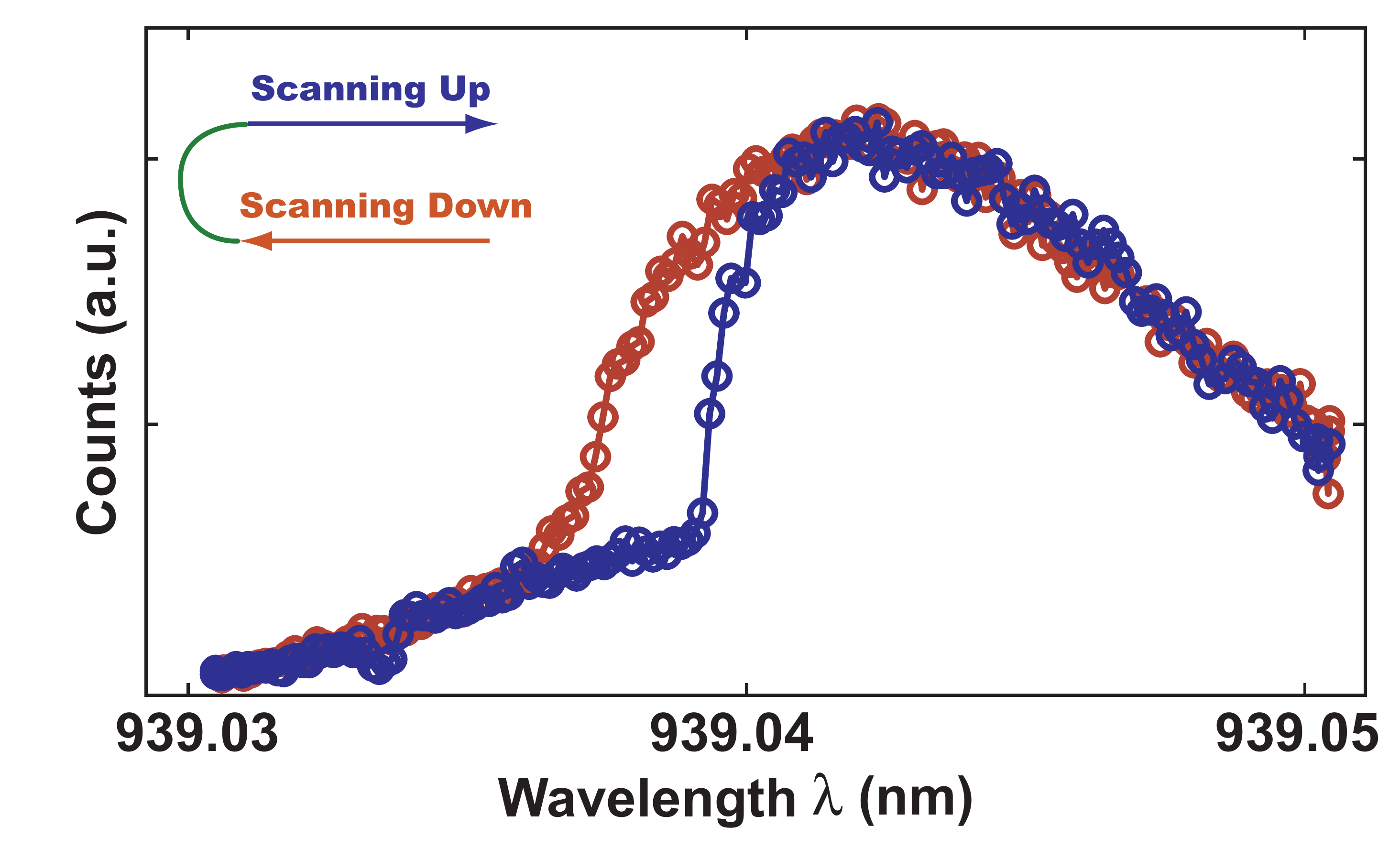}
\end{center}
\caption{Hysteretic absorption trace observed in a single-pulse experiment, while scanning the wavelength $\lambda$ of the pumping laser down and then up.}
\label{one-pulse-data}
\end{figure}
If observing the count rate as the wavelength of the optical pumping laser is scanned, one expects, in the absence of nuclear effects, a symmetric absorption lineshape.  This is not observed, however; instead, an asymmetric lineshape is observed.  Most notably, this lineshape shows hysteretic behavior, changing its shape depending on the direction of the scan and the history of the experiment.  This behavior is shown in \reffig{one-pulse-data}. Such effects were also observed in experiments is which the repumping was achieved with a second \CW-laser~\cite{sham_cpt,imamoglu_cpt}, rather than an ultrafast pulse as in our case.  Either way, a nuclear origin to the hysteresis is evident from the long timescale at which the hysteretic features survive.

\subsection{Free Induction Decay}
\begin{figure}[t]
\includegraphics[width=\columnwidth]{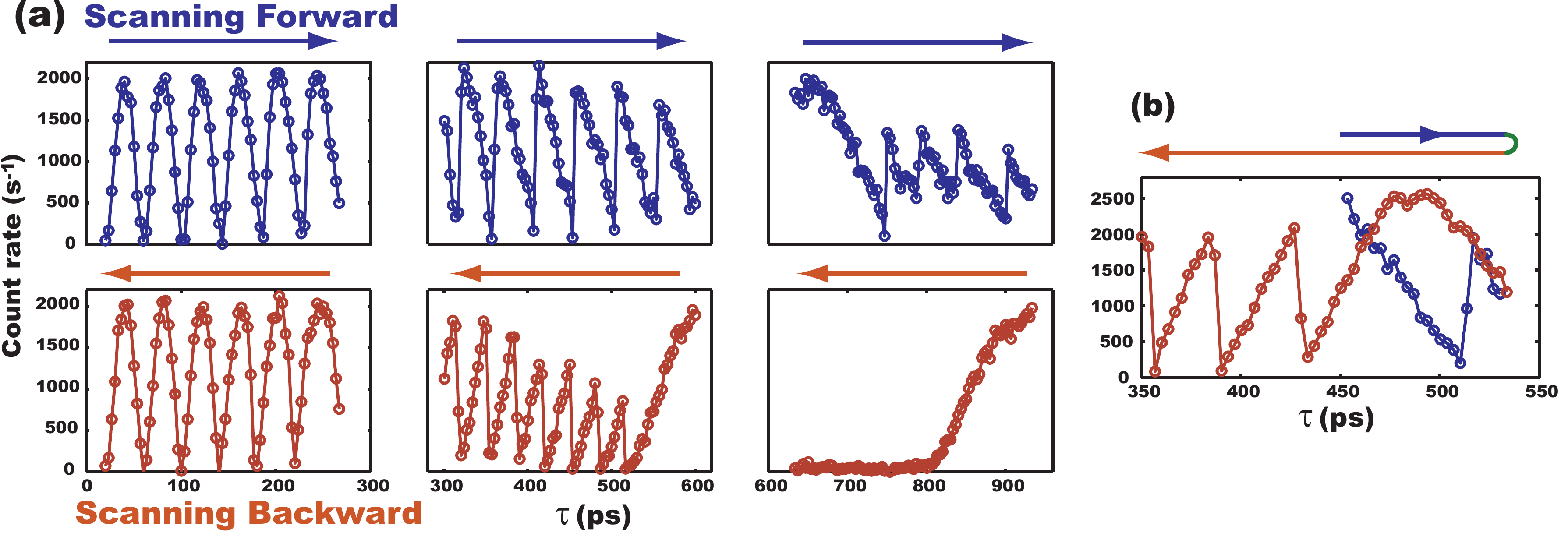}
\caption{(a) Experimental \FID\ count-rate
as a function of two-pulse time delay $\tau$.
(b) Experimental \FID\
fringe count-rate as $\tau$ is continuously scanned longer and then shorter, showing clear hysteresis.
}
\label{two-pulse-data}
\end{figure}

For the \FID\ experiment in the absence of any electron-nuclear spin feedback mechanisms, the nuclear spins would be expected to fluctuate randomly on a timescale slow compared to the Larmor precession, leading to random Overhauser shifts of the electron's Larmor frequency due to the contact-hyperfine interaction.  Repeated measurements of the same spin with different Larmor frequencies would lead to the sinusoidal \FID\ fringes decaying with a Gaussian shape on a timescale $T_2^*$, known to be on the order of nanoseconds from theoretical calculations. However, such a Gaussian decay was not observed.  Figure~\ref{two-pulse-data} shows the result of the \FID\ experiment. The top three traces show the fringes seen as the delay $\tau$ is increased, and the bottom three correspond to decreasing $\tau$.  The oscillatory fringes, rather than decaying,
evolve into a sawtooth pattern at high values of
$\tau$. Similar data were observed elsewhere as well:
see the top trace of Fig. 3b of Ref.~\onlinecite{kimarxiv}.  This pattern also shows hysteresis depending on
the direction in which $\tau$ is scanned. Fig.~\ref{two-pulse-data}(b)
illustrates the result of switching the scan direction.

\subsubsection{Spin echo}
\begin{figure}[t]
\begin{center}
\includegraphics[width=0.9\textwidth]{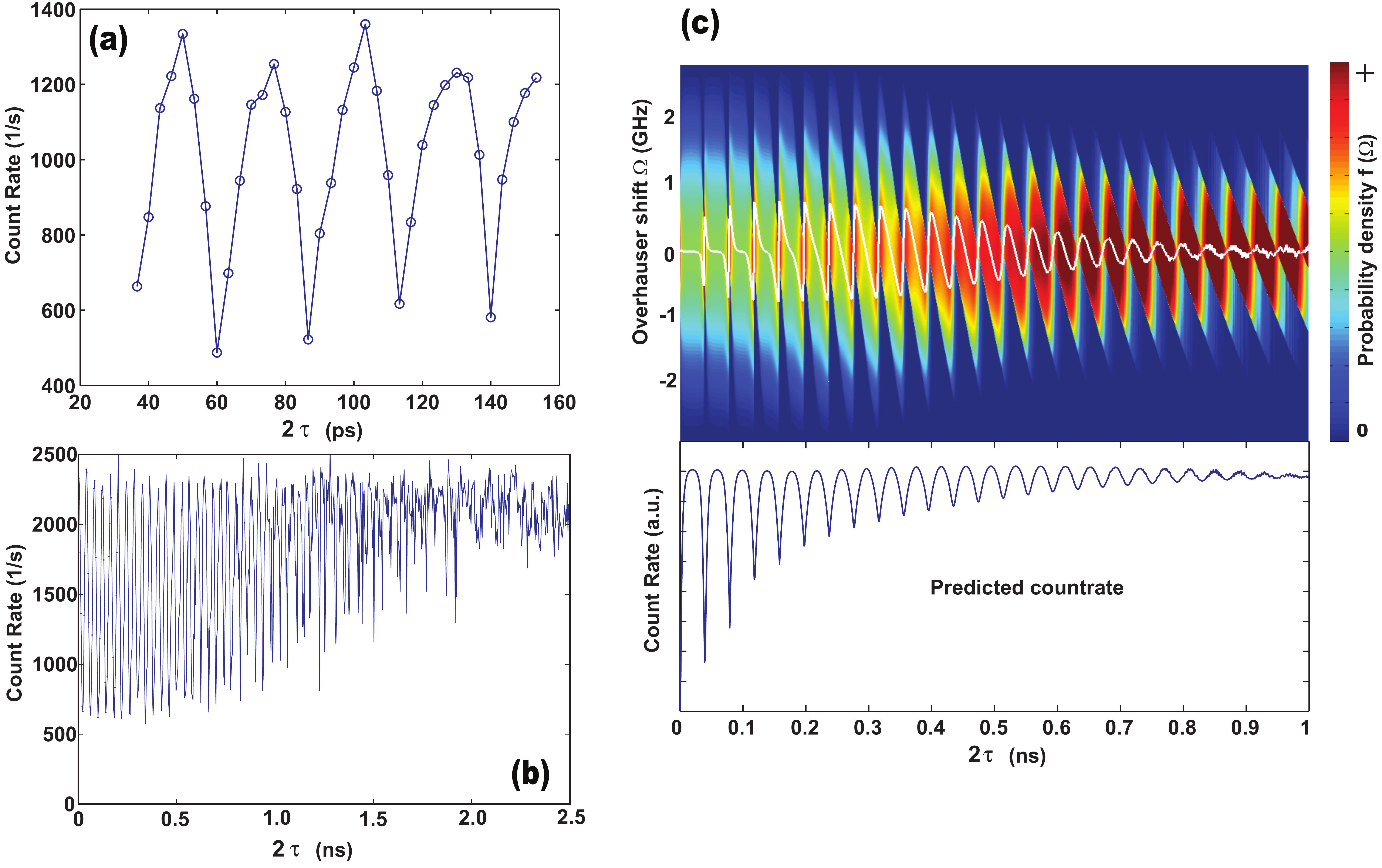}
\end{center}
\caption{
Experimental results for echo fringes and theoretical model.  (a) At low times $\tau$, ``spiky" fringes are observed.  These have clear deviation from sinusoidal behavior, with smoother curvature on the top of the oscillations and sharp cusps on the bottom.  This particular data was taken at magnetic field $B=3$~T with a $\pi$-pulse delay of $T=78$~ns.
(b)~At longer times $\tau$, the fringes are often observed to decay to a final count-rate which is not the expected ``middle" of the oscillations, but rather the maximum count-rate.  This particular data was taken at magnetic field $B=4$~T with a $\pi$-pulse delay of $T=13$~ns.
(c) Mathematical model explaining both spikey data and decay to a high value.  The top panel shows the $t\rightarrow\infty$ solution to \refeq{beq}, i.e. \refeq{infsol}, in the colorscale.  Superimposed in white is the average polarization $\langle\Omega\rangle$ for each $\tau$.  The resulting average countrate $\langle C_3(\Omega)\rangle$ is shown in the lower panel.  The timescale and Larmor frequency have been chosen not to match the data, but rather to easily show spiky oscillations and a decay to a high count-rate in a single plot. This particular model uses $D/\kappa=0.3$~ns$^{-2}$, $\ts\delta{e}/2\pi=25.3$~\GHz, $\alpha/\kappa=10^3$~ns$^{-2}$, and $\beta(\omega,\lambda_0)=\beta_0\exp(-\omega^2/2\sigma^2)$ with $\sigma/2\pi=$1.6~GHz and $\beta_0=0.1/\ts{T}{p}$.}
\label{three-pulse}
\end{figure}

The principal goal of the spin-echo measurement is to observe fringe visibility vanish as $T$ is increased, allowing a measurement of $T_2$. Such results are reported in Ref.~\onlinecite{pressecho}.  Not reported there are notable effects observed in the shape of the echo fringes as a function of $\tau$, which we qualitatively explain in the following sections.  A qualitative model will have to suffice, since the effects themselves depend on the sample, the sample history, and the averaging time, and hence can only be qualitatively reproduced.  Examples are shown in \reffig{three-pulse}(a,b).  The two principal effects are non-sinusoidal ``spikiness" of the echo fringes as a function of $\tau$ [\reffig{three-pulse}(a)], and a decay of the fringe count rate to a constant value substantially higher than the average count-rate [\reffig{three-pulse}(b)].  Neither effect can be explained from a simple inhomogeneous broadening model, and hence, in ensuing sections, we provide an explanation for these effects in terms of nuclear feedback effects.

\section{Physical Processes Driving Nuclear Polarization}
\label{physicalprocesses}

These data result from two competing processes, which ``push"
and ``pull" the nuclear polarization and therefore the
hyperfine (or Overhauser) shift in contrary
directions.  The ``push" process is due to trion excitation or decay, while the ``pull" process is due to nuclear spin diffusion.  In
what follows, we discuss the physical origins of these
processes and introduce a simple differential equation whose
stable solutions model our data well.

Let us begin by discussing the single-electron
ground state of the \QD, in which the electron spin,
effectively a dipole, sees primarily a simple Zeeman splitting
\be
\ham[e-Z] = \ts{g}{e}\mub B_0 S^z = \hbar\ts\delta{e} S^z.
\ee
Note that, in our convention, the (large) magnetic field is in
the $z$ direction.  The growth (and predominant uniaxial
strain) direction is taken to be the $x$ direction.  For the
electron spin, this strain plays little role, since a pure
$S=1/2$ dipole does not couple to the electric field gradients
introduced by strain.

The largest magnitude hyperfine interaction in a \QD\ is the Fermi contact hyperfine interaction,
\be
\ham[chf] = \sum_n A_n \vec{S}\cdot\vec{I}_n,
\ee
where $\vec{S}$ is the spin operator for the electron ($S=1/2$)
and $\vec{I}$ is the spin operator for the $n$th nucleus
($I=3/2$ for all isotopes in GaAs, and $I=9/2$ for the two
isotopes of In.)  Here the coupling constant $A_n$ varies
across the dot and by isotope; the sum $A=\sum_n A_n$ is on the order of 100~$\mu$eV.  The portion of this Hamiltonian proportional to $S^z$ is the dominant source of hyperfine shifts in a \QD; we denote this shift for some particular nuclear configuration as $\Omega = \sum_n A_n I^z_n/\hbar.$

Since nuclear spin-flip rates during trion decay are critical for our model, we must examine not the overall
strength of the hyperfine interaction, but rather the
probability of it flipping a nuclear spin during trion decay.  The lowest order nuclear-spin-flip rate occurs in second order
perturbation theory, which yields an estimate that, during
trion decay, the probability of a contact-hyperfine-induced
nuclear spin flip at nucleus $n$ is $|A_n/g\mub|^2,$ where
$g\mub$ is the electron Zeeman energy which must be compensated
for in the trion recombination process.  The Zeeman energy is
on the order of $30\GHz\times h$, putting this probability at
roughly $10^{-8}$ to $10^{-9}$ per nucleus, or with a timescale
of about 100~ms to 1~sec.  This process may play some role during trion emission, but to answer whether it is the dominant transition
or not, we must look at the other hyperfine interactions as well.

Before discussing the hole, we note that the \QD\ ground
state also features the electron-nuclear dipole-dipole
interaction,
\be
\ham[e-n] = \sum_n \frac{\mu_0}{4\pi}\hbar\gamma_n \ts{g}{e}\mub \biggl\langle
    \frac{3(\vec{r}_n\cdot\vec{S})(\vec{r}_n\cdot\vec{I}_n)-r_n^2(\vec{S}\cdot\vec{I}_n)}{r^5_n}
    \biggr\rangle,
\ee
where $\vec{r}_n=\vec{r}-\vec{R}_n$ for (assumed fixed) $n$th nuclear position $\vec{R}_n$ and the brackets $\langle\cdot\rangle$ now refer to an average
over the electron wavefunction. The integral over the \QD\
wavefunctions make this term very small for the $s$-like
symmetry of the ground state Bloch wavefunction of a conduction
electron close to the $\Gamma$-point in GaAs.  Certainly this interaction will be smaller than the contact term for flip-flop terms such as $S^- I^+$, and for these terms dipolar processes are clearly not competitive with the Fermi contact term.  This Hamiltonian
\emph{does} admit terms of the form $S^z I^x$, which
in-principle conserve the Zeeman energy.  Still, however, these terms are expected to be many orders of magnitude smaller than
the contact hyperfine term for $s$-shell conduction electrons.
The $S^z I^x$ term, for example, is proportional to
\be
\label{xz_term}
\left\langle \frac{\sin 2\theta\cos\phi}{r^3}\right\rangle,
\ee
which vanishes for a pure hydrogenic $s$-shell wavefunction and
nearly vanishes for a GaAs conduction electron.

The heavy hole of a trion is a different situation, however,
for two reasons. The first reason is that the strain-split $p$-like
configuration of the valence holes allows averages like
\refeq{xz_term} to survive.  The average of
$1/r^3$, where $r$ is the distance from a nucleus to a hole, is
a nontrivial number to calculate from first principles,
especially in the presence of strain.  For a back-of-the-envelope estimate, $\langle 1/r^3\rangle$ should be estimated as the size of a Wannier function on a single atomic site for the extended hole wavefunction.  More detailed calculations, such as that in Ref.~\onlinecite{coish_estimate}, put the strength of the heavy-hole nuclear dipole coupling in a GaAs quantum well or \QD, summed over all nuclei, in the
vicinity of 10~$\mu$eV. The average effective field of a heavy hole at an arsenic nucleus, for example, is therefore on the order of this number divided by $N\hbar\gamma_n$, where $N\sim 10^6$ is the number of nuclei and for Arsenic $\gamma_n\approx 5\times 10^7$(T sec)$^{-1}.$
Using these numbers, the local magnetic field seen by a nucleus
due to the dipole moment of the hole is on the order of 10~G.  We denote this field $B_{n}^{(\text{h})}.$

The second important reason the hyperfine interactions with the hole are different from the electrons is that light-hole/heavy-hole mixing plays a substantial role.  In particular, in addition to the Zeeman term, there is a strain Hamiltonian, combining to form a Hamiltonian of the form
\be
\ham[h] = a J_x^2 + b J_z,
\ee
for $J=3/2$.  The coefficient $a$ is proportional to the amount
of in-grown strain in the $x$-direction and $b$ is proportional
to the magnetic field.  In our sample, $a \gg b$.
Correspondingly, what we generally refer to as a heavy-hole trion is, at these fields, a combination of a heavy and light hole, and
its dominant spin direction in the ground state is at some
angle between the $x$ and $z$, closer to the $x$-direction.
Suppose this direction is labeled $\hat{r}_0$; then the dipolar
coupling between the hole and nucleus $n$ contains the
hole-Zeeman-energy-conserving term $(\vec{J}\cdot\hat{r}_0)
I^x$, which will in general not vanish in this low symmetry
situation.  In fact, as we discuss below, this term is only a
bit more than an order of magnitude weaker than the Fermi
contact term. Crucially, though, the $(\vec{J}\cdot\hat{r}_0)
I^x$ term is nearly energy-conserving, and therefore in
second-order perturbation theory, the energy denominator is
very small, on the order of the nuclear Zeeman energy of a few
MHz instead of the electron Zeeman energy of a few GHz.  As a
result, the overall transition rate during trion emission is at
least an order of magnitude larger than that due to the contact
hyperfine interaction.  Fermi's golden rule in second
order allows an estimate of the rate $\Gamma_n$ at which a
trion hole nearly polarized along the sample growth axis
(orthogonal to the magnetic field) randomly flips the $n$th
nuclear spin during spontaneous emission, with the nuclear Zeeman energy compensated by the broad width of the emitted photon
($\gamma/2\pi\sim 0.1$~GHz) and the photonic
density of states negligibly changed by the Zeeman energy of
the nucleus.
The result is $\Gamma_n \approx
(B_{n}^{(\text{h})}/B_0)^2\gamma \approx 1/(10\ms)$.

Although we conclude that hole-hyperfine interactions are responsible for the nuclear polarization effects in our experiment, this conclusion should not be construed to mean that electron hyperfine interactions play no role at all, or never affect nuclear spin dynamics in \QDs.  They do and certainly can, especially in experiments in the Faraday geometry where the strain and field directions are parallel, and the hole dipolar interaction plays a lesser role. However, we believe that the particular phenomena we report here are most likely explained by the hole dipolar interaction. The conclusion that hole dipolar interactions contribute predominantly to nuclear feedback effects in optically controlled InGaAs \QDs\ was also reached in Ref.~\onlinecite{sham_cpt}; more detailed discussion appears in Ref.~\onlinecite{sham_cpt_preprint}.

The ``pull" process of our model is the presence of nuclear spin diffusion,
a process known to limit nuclear polarization rates in \QDs~\cite{gammon_study}.  By ``nuclear spin diffusion," we are referring to the motion of
nuclear spin polarization due to the nuclear dipole-dipole
interaction.  Technically speaking, diffusion in the most classical sense is not a complete description of the loss of nuclear polarization in a \QD.  The reason is that the hyperfine field of the electron introduces a magnetic
field gradient, which makes dipole-dipole induced changes in
the nuclear field energy-nonconserving.  Polarization may still
move and change, however, due to the finite dipolar width in
the large three-dimensional bath of spins~\cite{gr73} and due to
fluctuations of the gradient~\cite{gammon_study}.
In general, the magnitude of the nuclear dipole-dipole coupling sets a spin-diffusion timescale on the order of milliseconds.  However, the more complex processes occurring in the presence of the hyperfine field gradient allow for a hierarchy of timescales. The fastest,  millisecond timescales would describe nuclei approaching a quasi-equilibrium dipolar configuration, which may depend on the history of magnetic interactions inside the dot.  We denote the timescale to reach this quasi-equilibrium $\kappa^{-1}$.
Eventually, diffusion should cause nuclear polarization to ``leak out" of the \QD\ and reach thermal equilibrium.  This timescale may be far longer, however, on the order the nuclear $T_1$ rate, which is observed to be on the order of 20~minutes in our system.

We therefore see that our experiment features five very distinct timescales, and our model will depend crucially on the assumption that they remain distinct.  The fastest timescale is the dynamics during picosecond pulses, which play no role for the nuclei; hence the electron rotations are considered instantaneous.  The next slower timescale is that of the pulse
sequence and resulting electron-spin dynamics, repeated
continuously with a repetition period on the order of hundreds of nanoseconds in the one- and two-pulse experiments, and on the order of $2T$ in the spin-echo experiments.  This is still much faster than the third timescale, that of the millisecond-speed nuclear dynamics.  Next, the averaging timescale of the measurement is much longer, on the order of several seconds, allowing the nuclei substantial time to reach quasi-equilibrium.  The longest timescale, that of nuclear $T_1$, is far longer still, and plays no role in our model.

\section{Mathematical Model}
\label{themodel}

The previous section motivated the theory that trion creation or decay drives nuclear polarization effects in our system.  This theory provides no bias for which direction that polarization is driven, however.  In our model, trion transitions drive an unbiased random walk in nuclear
polarization, with a \emph{rate} proportional to the
probability that a trion is created by the pulse
sequence.

Our model therefore is based on a trion-driven diffusion equation.  We describe the nuclear hyperfine shift as a random
variable $\Omega$ with time-dependent probability density function (\pdf) $f(\Omega,t)$.  Time-dependent averages over this \pdf\ are given by $\langle g(\Omega) \rangle = \int d\Omega g(\Omega)
f(\Omega,t)$, and in particular we write $\omega(t)=\langle\Omega\rangle$.

\begin{figure}
\begin{center}
\includegraphics[width=0.9\textwidth]{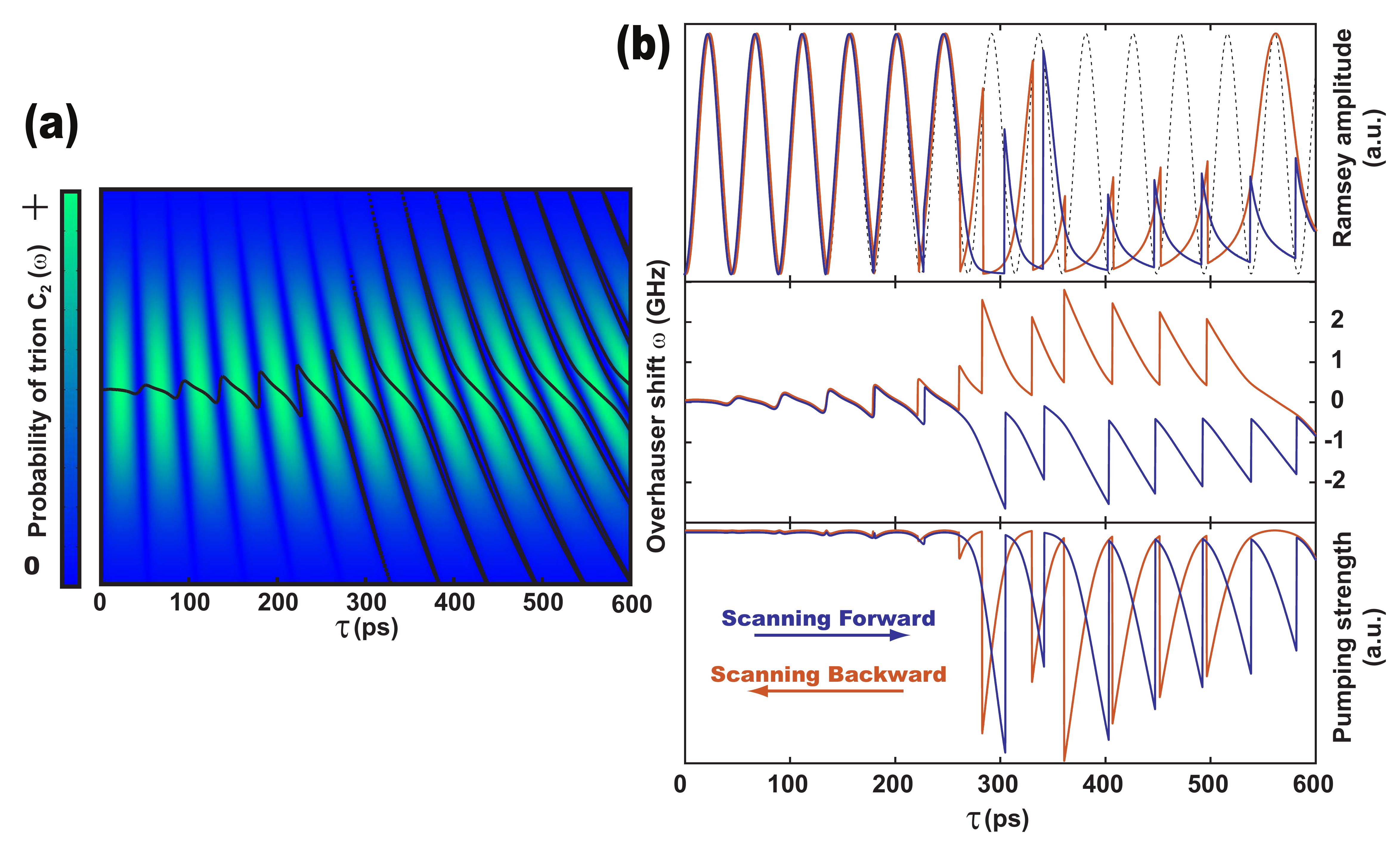}
\end{center}
\caption{
(a) The colorscale shows the count rate $C_2(\omega)$ as a function of average Overhauser shift $\omega$ (vertical axis, scale shown in GHz in second panel of subfigure 2), and two-pulse delay $\tau$ (horizontal axis).  The green areas indicate where a higher count-rate is expected. Oscillations in the horizontal directions at frequency $\ts\delta{e}+\omega$ are due to Ramsey interference; the Gaussian envelope in the vertical
direction is due to the reduction of optical pumping with detuning. The superimposed black line indicates stable points where $\partial\omega/\partial t=0$ according
to \protect\refeq{de}.
(b) Results of the two-pulse model, to be compared against \reffig{two-pulse-data}.  The top panel shows the predicted shape of the Ramsey fringes.  These result from an Overhauser shift shown on the second panel, which hysteretically follows stable regions of the black curve in (a).  The third panel shows the pumping efficiency at each $\tau$.  This particular model uses $\kappa/\alpha=10^4$~ps$^2$ and $\beta(\omega,\lambda_0)=\beta_0\exp(-\omega^2/2\sigma^2)$ with $\sigma/2\pi=$1.6~GHz and $\beta_0=3/\ts{T}{p}$.}
\label{two-pulse-model}
\end{figure}

The trion creation probability will be the crucial term in our diffusion equation.  We derive this function as follows.  Let us suppose that, for any pulse sequence, the average value of the electron spin for each cycle before optical pumping is \Sb, and the average spin afterwards is \Sa.  Note that coherences are not important for the dynamics studied here, so $S$ refers only to $S^z$, and the $z$ is dropped.  During the optical pumping, the spin is driven to a final equilibrium population $\ts{S}{p}^z$, as shown in \reffig{sequences}, following a standard Bloch equation
\be
\frac{d}{dt}\langle S \rangle = -\beta(\Omega,\lambda) [\langle S \rangle-\ts{S}{p}^z],
\ee
where $\beta(\Omega,\lambda)$ is a pumping rate, which depends on the Overhauser shift $\Omega$ and the pumping wavelength $\lambda$.  On resonance, this rate has been measured to be on the order of ns$^{-1}$ in our system~\cite{pressnature}.  After evolving for a time $\ts{T}{p}$, we have
\be
\Sa = \ts{S}{p}^z(1-e^{-\beta(\Omega,\lambda)\ts{T}{p}})+\Sb e^{-\beta(\Omega,\lambda)\ts{T}{p}}.
\ee
This relation may then be combined with a second relation between $\Sa$ and $\Sb$ that depends on the pulse sequence.  In the single-pulse case, the $\pi$ pulse assures $\Sa=-\Sb$, and therefore the one-pulse trion count rate is
\be
C_1(\Omega)=\Sa-\Sb=2\ts{S}{p}^z\tanh[\beta(\Omega,\lambda)\ts{T}{p}/2].
\ee
Note that $C_1(\Omega)$ depends on $\lambda$, but we shall not explicitly notate this dependence for brevity. For the two-pulse experiment, the pumping laser wavelength is held fixed at $\lambda=\lambda_0$, but the varying pulse separation of $\tau$ gives
\be
\Sb = \cos[\phi_0+(\ts\delta{e}+\Omega)\tau]\Sa.
\ee
The phase $\phi_0$ depends on offsets in the zero of $\tau$.  Strictly speaking, at $\tau=0$, the two pulses overlap and $\phi_0=\pi$.  In practice, there is a small offset in $\tau$ that plays little role in the important features of the dynamics.  For a more direct comparison to \reffig{two-pulse-data}, we set $\phi_0=0$.  Hence in the two-pulse case the average number of trions created per cycle is now given by
\be
\label{countrate}
C_2(\Omega)=\Sa-\Sb=\ts{S}{p}^z
    \frac{[1-\exp(-\beta(\Omega,\lambda_0)T)]
          [1-\cos[(\ts\delta{e}+\Omega)\tau]]}
    {1-\cos[(\ts\delta{e}+\Omega)\tau]\exp(-\beta(\Omega,\lambda_0)T)}.
\ee
Now $C_2(\Omega)$ depends on $\tau$, although again we will not notate this explicitly.  This trion creation probability is plotted in the green and blue colorscale of \reffig{two-pulse-model}(a) as a function of pulse delay $\tau$ and Overhauser shift $\Omega$.  This function oscillates sinusoidally with increasing $\tau$ due to the spin's Larmor precession, and would lead to \FID\ oscillations for fixed $\Omega$. The Overhauser shift $\Omega$ changes the frequency of Larmor precession.  The trion creation probability drops to 0 for large values of $|\Omega|$ because the trion transition shifts away from resonance with the optical pumping laser, leading to reduced pumping efficiency ($\beta(\Omega,\lambda_0) \rightarrow 0$) and therefore reduced trion creation.




The three-pulse case, at fixed $\lambda=\lambda_0$ and fixed $T$, gives a very similar result.  The only difference is that the relative position of the $\pi$ pulse drives the rotation angle, rather than the second $\pi/2$ pulse, hence effectively doubling $\tau$:
\be
C_3(\Omega,\tau)=C_2(\Omega,2\tau).
\ee

The diffusive terms of our equation are generated by the following argument, supposing a nuclear flip causes a change $d\Omega$ in the Overhauser shift in time $dt$.  The probability of finding an Overhauser shift between $\Omega$ and $\Omega+d\Omega$ at time $t+dt$ is, by definition, $f(\Omega,t+dt)dx$, and is the sum of three possibilities.  In time $dt$, $\Omega$ may increase by $d\Omega$ from $\Omega-d\Omega$, it may decrease by $d\Omega$ from $\Omega+d\Omega$, or it may stay the same at $\Omega$.  If $\Omega$ jumps up by $d\Omega$ with probability $[\delta+\epsilon
C_j(x)]/2$, where $\delta$ is a trion-independent spin diffusion probability and $\epsilon$ quantifies the total trion-nuclear
spin-flip probability, and if $\Omega$ jumps down with the same probability,
then the probability density of seeing $\Omega$ at time $t+dt$ is
\be
f(\Omega,t+dt)=\frac{1}{2}[\delta+\epsilon C_j(\Omega+d\Omega)] f(\Omega+d\Omega,t) +
\frac{1}{2} [\delta+\epsilon C_j(\Omega-d\Omega)]f(\Omega-d\Omega,t) + [1-\delta-\epsilon C_j(\Omega)]
f(\Omega,t).
\ee
In the limit that $dt$ and $d\Omega$ become vanishingly small, and defining
$\alpha = d\Omega^2\epsilon/2dt$ and $D=d\Omega^2\delta/2dt$, this expression becomes
\be
\frac{\partial}{\partial t}f(x,t) = D \frac{\partial}{\partial\Omega^2} f(\Omega,t) +\alpha
\frac{\partial}{\partial\Omega}\biggl[C_j(\Omega)
\frac{\partial}{\partial\Omega}f(\Omega,t)\biggr].
\ee
The parameter $\alpha$
may be estimated as $\sum_n \Gamma_n (A_n/\hbar)^2 \approx
\ts\delta{e}^2/1$~sec.

An additional term is required to describe the phenomena by which nuclear spin diffusion reduces large shifts and results in a finite-width distribution (of nominal width of order $1/T_2^*$.)  These effects may be phenomenologically captured by a term which drives the mean of $f(\Omega,t)$ to zero at a rate $\kappa$.
The entire evolution for $f(x,t)$ might therefore considered to be
\be
\label{beq}
\frac{\partial}{\partial t} f(\Omega,t) = \kappa
\frac{\partial}{\partial\Omega}[\Omega f(\Omega,t)] +
\frac{\partial}{\partial\Omega}
\biggl\{\left[D+\alpha C_j(\Omega)\right]\frac{\partial}{\partial \Omega}f(\Omega,t)\biggr\}.
\ee
Note that if $f(\Omega,t)$ is a proper \textsc{pdf} with unit integral, then this equation conserves that property.

Before proceeding, let us note that an equation similar to \refeq{beq} has been used to describe nuclear feedback effects in optically controlled \QD\ experiments previously.
Reference~\onlinecite{greilich_nuclear} derives, as Eq.~(16) of the
Supporting Online Material, a diffusion equation similar to the
above.  However, this experiment, as well as that in Ref.~\onlinecite{Reinecke_nuclear}, are quite different from the one we describe here.  That experiment works with an inhomogeneous ensemble rather than a single dot, and employs a very different pulse sequence; we have a \CW\ optical pumping field
which is not present in those other experiments.  This pumping field is likely (although not provably) the dominant reason the model we present in our manuscript is successful while attempts to apply the models in Refs.~\onlinecite{greilich_nuclear,Reinecke_nuclear} have
failed in our case.  A key assumption in Refs.~\onlinecite{greilich_nuclear,Reinecke_nuclear} is that the asymmetry in the direction of the average electron spin direction during the pulse sequence, or the ``light-induced decoherence" of the ground-state electron induced by the pulse-modulated electron hyperfine interaction, are key factors in the feedback process. Originally, these references led us to believe that this would be true in our experiment as well.  However, in our experiment, we pump the electron to a particular spin state after each rotation pulse pair, an element not present in the experiments described by Refs.~\onlinecite{greilich_nuclear,Reinecke_nuclear}.
As a result, the average electron polarization in our experiment is almost constant with respect to the parameter $\tau$, and therefore does not affect our data, and decoherence of electron spin precession is far less important, since only during a very small fraction of time in our one- and two-pulse sequences is the electron actually precessing (in contrast to
Refs.~\onlinecite{greilich_nuclear,Reinecke_nuclear} where the electrons are precessing most of the time).  We have substantial control over our pulse sequence, and we therefore tested this assumption by varying the spacing between pulses, the duration in optical pumping, etc.  Doing so allowed us to change the effective electron spin orientation during the bulk of the experiment and the relative degree of ``light-induced
decoherence" and we discovered that these factors could not
explain our data.  In total, our experimental explorations led to the following critical conclusion: \emph{Only the rate of real trion creation during optical pumping is important for nuclear effects in our experiment, {\bf\it independent} of the average electron polarization,} unlike in Refs.~\onlinecite{greilich_nuclear,Reinecke_nuclear}. (Of course, in a single experiment, the electron polarization right before the optical pumping step and the trion creation rate are correlated.)


We now examine \refeq{beq} in three important regimes.  The first regime is where trion-induced effects are unimportant.  One way to achieve this is to set $\alpha\ll D$; in that case the steady-state solution is a Gaussian distribution with mean decaying to zero at rate $\kappa$ and variance which exponentially approaches $D/\kappa$ in time.  In this respect, we might associate $D/\kappa$ with $1/T_2^*$, so that this equation could describe a simple picture of the \FID; that is, the fluctuation rate would be independent of
the pulse-sequence and the fringes of $\langle C(\Omega,\tau)\rangle$ would undergo Gaussian decay with time-constant $T_2^*$.  Another regime in which simple Gaussian decay is seen is if $f(\Omega,0)$ is Gaussian, and the amount of time $t$ the system is allowed to evolve is small.  In this case, solutions remain Gaussian and ordinary dephasing processes are observed.  This regime is accessed only in experiments employing a spin-echo.  Critically, in a spin-echo experiment there is substantial time for the nuclei to diffuse unhindered by electron polarization, since for the majority of the pulse sequence, the electron spin is precessing in its Bloch sphere, which effectively decouples the electron from the nuclear dipole-dipole dynamics.  The subsequent lack of $z$-oriented polarization results in a zero average hyperfine gradient on the nuclear spins, allowing maximal diffusion and therefore the $D \gg \alpha$ regime.  If in addition relatively low averaging times are employed (resulting in decay dominated by observable noise), trion effects may be ignored, Gaussian decay may be observed,  and $T_2^*$ may be measured, as performed in Ref.~\onlinecite{pressecho}.

A second regime is where $\alpha \sim D$, and a steady-state solution is reached ($t\rightarrow\infty$).  This is believed to occur when diffusion is relatively unhindered, as in a spin-echo experiment, but long averaging times are employed.  In this regime, this flux-conservative PDE is well-behaved, and predicts that $f(\Omega,t)$ eventually evolves, from any initial condition, into a steady-state solution.  That steady-state solution may be written
\be
\label{infsol}
f(\Omega,t\rightarrow\infty) = \\
f(0,t\rightarrow\infty) \exp\biggl[-\kappa \int_0^\Omega \frac{u}{D+\alpha C_j(u)}du\biggr].
\ee
The prefactor $f(0,t\rightarrow\infty)$ may be found via normalization.  This solution describes a distribution with a mean that fluctuates around 0 as $\tau$ is increased, and whose variance varies periodically with $\tau$.  Such a solution is plotted in \reffig{three-pulse}(c), upper panel.  If this solution is used to find the average expected count rate in a spin-echo experiment, $\langle C_3(\Omega)\rangle$ as a function of $\tau$, the result shows qualitative features observed in spin-echo experiments, as seen in \reffig{three-pulse}(c), lower panel.  Oscillations at short times show spiky, non-sinusoidal character, and when the oscillations damp at high values of $\tau$, they do so to a value close to the maximum count-rate.  The physical origin of these effects in this model is evidently that, at those places where the fringes look spiky, the nuclei have become polarized and their distribution is substantially narrowed.  As $\tau$ is increased, this degree of narrowing decreases at the spiky (minimum count rate) locations, but there is little change in the distribution at the maximum count rate locations, accounting for the decay to near the the maximum count rate.

While this solution captures some of the puzzling features of spin-echo fringes, it cannot be a complete model for the \FID\ data shown in \reffig{two-pulse-data}, or even the hysteresis in \reffig{one-pulse-data}.  Indeed, a linear PDE such as \refeq{beq} under assumptions of full equilibrium cannot show hysteresis.  However, only one more simple assumption is needed to arrive at a model which does match our data well.  This final assumption is that the width of $f(\Omega,\tau)$ is constrained to be small; much smaller than the inverse Larmor period, despite the influence of trion-induced diffusion.  This is possible if the nuclear system has neither the time nor the freedom to fully fill the distribution predicted by \refeq{beq}, possibly because nuclear spin diffusion is suppressed by hyperfine magnetic field gradients from polarized electron spins.  Recall that in our \textsc{fid} experiment, the electron spends the bulk of its time with some polarization on the $z$-axis.  This assumption is further motivated by the fact that, in \textsc{fid} experiments, the sawtooth-like fringes we observe last for at least an order-of-magnitude longer than $T_2^*$, suggesting that the distribution $f(\Omega,t)$ is substantially narrowed.

Under this assumption of a narrow distribution due to suppressed diffusion, we model only the first moment of $f(\Omega,t)$, $\omega(t)=\langle \Omega(t) \rangle$, and assume the second moment is much smaller than $1/\tau$.  Using integration by parts,
\be
\label{halfway}
\frac{\partial\omega}{\partial t} = -\kappa \omega +
\alpha\left\langle \frac{\partial C_j}{\partial \Omega}
\right\rangle.
\ee
Equation~(\ref{halfway}) is not a closed system, because it still requires knowledge of all moments of $f(\Omega,t)$ to complete the average over $C_j'(\Omega)$.  However, to approximately remove the average, we apply our assumption about the width of $f(\Omega,t)$ to write
\be
C_j'[\Omega]\approx C_j'[\omega(t)]+[\Omega-\omega(t)]C_j''[\omega(t)]+\cdots,
\ee
leaving, to lowest order,
\be
\frac{\partial\omega}{\partial t} = -\kappa \omega +
\alpha C_j'(\omega).
\label{de}
\ee
Invoking our separation of timescales, we presume $\omega(t)$
evolves from its initial value (set by the last chosen value of $\lambda$ or $\tau$) to a quasi-equilibrium final value $\ts\omega{f}$. This final value determines the expected count rate $C_j(\ts{\omega}{f})$ at this value of $\lambda$ or $\tau$.  We solve by
assuming $\omega(0)=0$ at the first attempted value of $\lambda$ or $\tau$, and then we scan the parameter as in the experiment, finding the steady-state solution of \refeq{de} at each value.

\begin{figure}
\begin{center}
\includegraphics[width=0.6\textwidth]{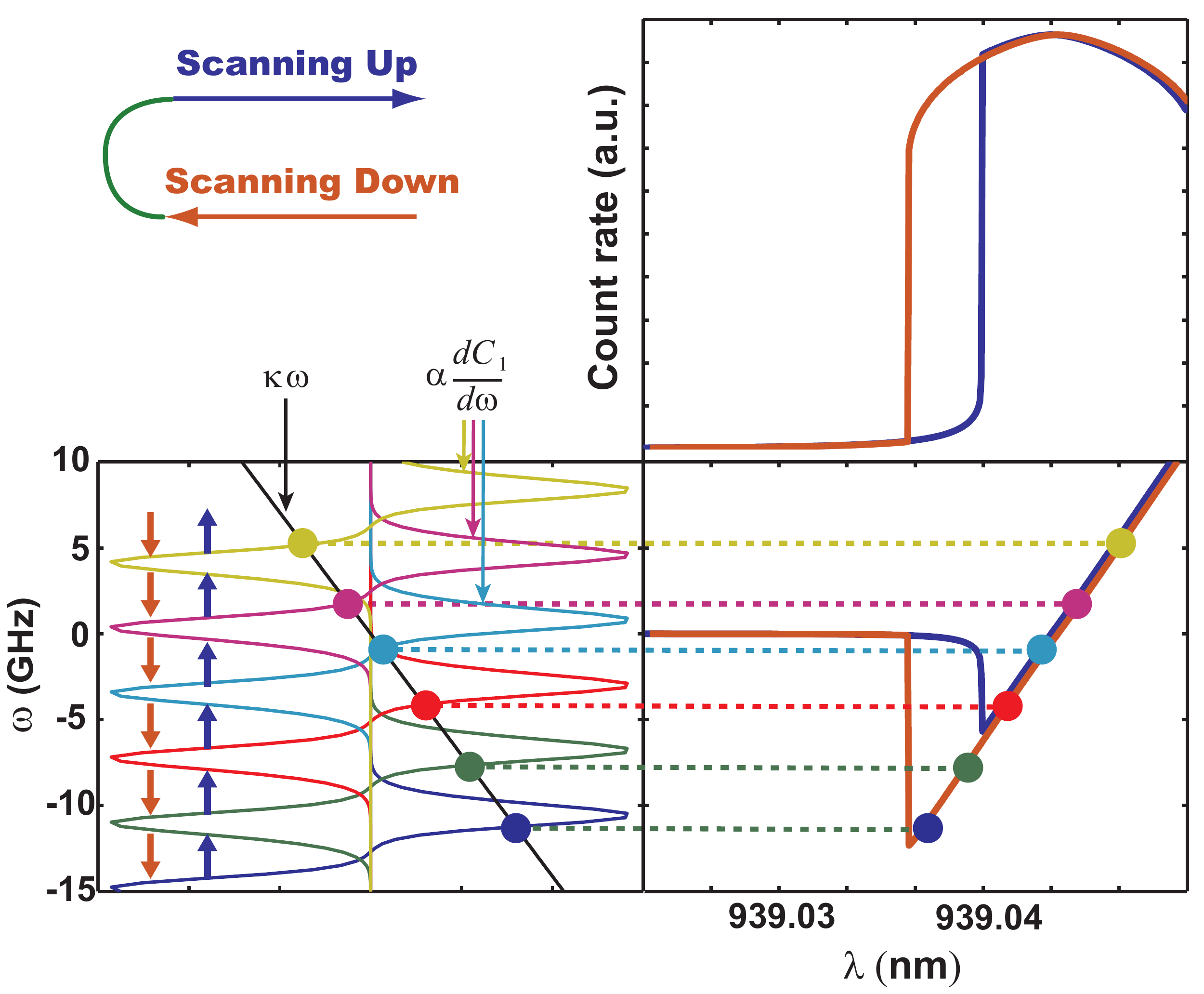}
\end{center}
\caption{Elements of the model for the one-pulse experiment.  The panel on the lower left shows $\kappa\omega$ (black line) and $\alpha dC_1(\omega)/d\omega$ for several values of $\lambda$, as $\lambda$ is scanned downwards (indicated by the brown arrows) and then upwards (indicated by the blue arrows).  The circles indicate locations where $\kappa\omega=\alpha dC_1(\omega)/d\omega$, i.e. the steady-state solutions of \refeq{de} for $j=1$.  As $\lambda$ is scanned downward, the polarization $\omega$ ``surfs" the same edge of the of the $\alpha dC_1(\omega)/d\omega$ curve, and $\omega$ steadily decreases, as shown in the lower-right panel.  Eventually, at low enough $\lambda$, the peak of $\alpha dC_1(\omega)/d\omega$ is passed and the only remaining steady-state solution shifts to $\omega=0$.  This solution is maintained as $\lambda$ is scanned upwards, until the peak of $\alpha dC_1(\omega)/d\omega$ reaches $\omega=0$, at which point $\omega$ again surfs the $\alpha dC_1(\omega)/d\omega$.  The resulting countrate, $C_1(\omega)$, is shown in the upper right panel, and is to be qualitatively compared to \reffig{one-pulse-data}.  This particular model uses $\kappa/\alpha=3\times{10}^4$~ps$^2$,}
\label{one-pulse-model}
\end{figure}

The nonlinearity in this ordinary differential equation now allows hysteresis.  The hysteresis is most easily understood in the one-pulse experiment, whose model is illustrated in \reffig{one-pulse-model}.  Scanning the parameter $\lambda$ is easy to envision graphically, since for this sequence a change in scanning frequency $\delta\omega = \delta(2\pi c/\lambda)$ amounts to a linear shift in $C_1'(\omega)$, i.e.  $C'_1(\omega)\rightarrow C'_1(\omega+\delta\omega)$.  The steady-state solution is where $\alpha C'_1(\omega)=\kappa\omega$; several such solutions exist at particular values of $\lambda$.  When one such solution lives on the edge of a peak of $C'_1(\omega)$, it ``surfs" that edge until its crest.  This picture explains bistability and hence hysteresis; see \reffig{one-pulse-model}.  With the exception of experimental differences, this model is similar to that used to explain hysteresis in \CW\ experiments~\cite{sham_cpt}.

This basic model allows an analysis of the more complicated two-pulse experiment.  The essence of our model is that the  system quasi-equilibrates to a stable value of nuclear polarization that lives on the edge of the fringes shown in \reffig{two-pulse-model}(a); with our separation of timescales, the nuclear polarization then ``surfs" along the edge of this function as $\tau$ is changed.  As $\tau$ is
increased, $|\omega|$ increases causing the observable photon
count to decrease due to the reduced degree of optical pumping. When $|\omega|$ is so high that pumping is ineffective
($\beta(\omega)\rightarrow 0$) and the trion-induced walk
stops, spin-diffusion causes the system to drift back to a new
stable magnetization at a lower value of $|\omega|$, and the
process continues.

Figure~\ref{two-pulse-model} shows the modeled
$C_2(\ts{\omega}{f})$ and $\ts\omega{f}/2\pi$ as a function
of $\tau$.  This particular model used
$\kappa/\alpha=10^4$~ps$^{2}$, which provides the estimate
$\kappa^{-1}\sim 100$~ms, a relevant timescale for
dipolar-induced nuclear spin diffusion.  This choice of
parameters reproduces the qualitative shape of the data quite
well, and quantitatively reproduces the location where
sinusoidal fringes evolve into sawtooth-like fringes.
Detailed curve fitting is
not reasonable for this process, since the quantitative
behavior depends on a random initial conditions.
Qualitative differences are dominated by the random conditions
that develop when the stage is moved on its rail, forming the
breaks between data sets in \reffig{two-pulse-data}(a).

Despite the small differences, the success of our simple mathematical model in qualitatively reproducing the initially mysterious hysteretic and non-sinusoidal phenomena in one-, two-, and three-pulse experiments provides high confidence that the basic physics of nuclear feedback is understood in this system.

\section{Conclusion}

The effects we explain here, along with similar effects reported elsewhere~\cite{vandersypen_singlet-triplet,Reilly_Zamboni_Science,delftlock,new_yacoby,greilich_nuclear,Reinecke_nuclear}, may be useful for future coherent technologies employing \QDs.   Both the two- and three-pulse sequences may serve as
``preparation steps" for a qubit to be used in a quantum
information processor, as it tunes the qubit to a master
oscillator.  The efficacy of this tuning is evident from the reduced width of $f(\Omega,t)$ predicted by our model, and appears experimentally as spiky fringes in spin-echo experiments and saw-tooth fringes in \FID\ experiments.
Tuning to a master oscillator is critical for coherent qubit control~\cite{fast_rotations_prop}.   In the shorter term,
the phase-controlled $\delta$-function-like rotation pulses we employ are very promising for strong dynamical
decoupling (\abbrev{DD})~\cite{lidar08}.   Unfortunately
many \abbrev{DD} schemes, especially those that compensate for
pulse errors such as the workhorse Carr-Purcell-Meiboom-Gill (\abbrev{CPMG})
sequence,
require some method to tune the \QD's Larmor period
to an appropriate division of the pulse-separation time.  Judicious use of the effects we have modeled here may enable such experiments in self-assembled \QDs\ in the near future.

In summary, we have observed nonlinear nuclear feedback
effects in a single charged \QD\ resulting from the countering
processes of random nuclear walks driven by trion creation, the finite width of optical absorption, Overhauser-shifted Larmor
precession, and nuclear spin diffusion.
This feedback may be
employed for tuning the electron Larmor period to a particular
pulse separation time for more complex pulses sequences.
Future work might involve extensions of this theory to explain non-Markovian effects of nuclei under more complex pulse sequences, as well as exploiting them to extend \QD-based quantum memories.

\section{Acknowledgements}

We thank Erwin Hahn for valuable discussions. This work was
supported by \textsc{NICT, MEXT, NSF CCF0829694}, the State of
Bavaria, and Special Coordination Funds for Promoting Science
and Technology. PLM was supported by the David Cheriton
Stanford Graduate Fellowship.

\bibliographystyle{spiebib}

\end{document}